\documentclass[journal=jctcee,manuscript=letter]{achemso}
\setkeys{acs}{maxauthors=4,etalmode=truncate,chaptertitle =true,articletitle=true}
\newcommand{\onlinecite}[1]{\hspace{-1 ex} \nocite{#1}\citenum{#1}} 
\usepackage{amssymb}
\usepackage{amsmath}
\usepackage[usenames,dvipsnames]{xcolor}
\usepackage{graphicx}
\usepackage{float}
\usepackage[normalem]{ulem}
\usepackage{subcaption}
\usepackage{natbib}
\usepackage{epstopdf}

\newcommand{\ket}[1]{\left|#1\right\rangle}

\usepackage{xcolor}

\usepackage{multirow}
\usepackage{bm}

\title{Highly Accurate Prediction of Core Spectra of Molecules at Density Functional Theory Cost: Attaining sub eV Error from a Restricted Open-Shell Kohn-Sham Approach.}

\author{Diptarka Hait}
\email{diptarka@berkeley.edu}
\affiliation
{{Kenneth S. Pitzer Center for Theoretical Chemistry, Department of Chemistry, University of California, Berkeley, California 94720, USA}}
\alsoaffiliation{Chemical Sciences Division, Lawrence Berkeley National Laboratory, Berkeley, California 94720, USA}
\author{Martin Head-Gordon}
\email{mhg@cchem.berkeley.edu}
\affiliation
{{Kenneth S. Pitzer Center for Theoretical Chemistry, Department of Chemistry, University of California, Berkeley, California 94720, USA}}
\alsoaffiliation{Chemical Sciences Division, Lawrence Berkeley National Laboratory, Berkeley, California 94720, USA}

\begin{document}
	\begin{abstract}
 We present the use of the recently developed Square Gradient Minimization (SGM) algorithm for excited state orbital optimization, to obtain spin-pure Restricted Open-Shell Kohn-Sham (ROKS) energies for core excited states of molecules. \textcolor{black}{The SGM algorithm is robust against variational collapse, and offers a reliable route to converging orbitals for target excited states at only 2-3 times the cost of ground state orbital optimization (per iteration)}. ROKS\textcolor{black}{/SGM} with the modern SCAN/$\omega$B97X-V functionals is found to predict the K edge of C,N,O and F to a root mean squared error of $\sim$0.3 eV. \textcolor{black}{ROKS/SGM is equally effective at predicting L edge spectra of third period elements, provided a perturbative spin-orbit correction is employed}. This high accuracy can be contrasted with traditional TDDFT, which typically has greater than $10$ eV error and requires translation of computed spectra to align with experiment. ROKS is computationally affordable (having the same scaling as ground state DFT, and a slightly larger prefactor) and can be applied to geometry optimizations/ab-initio molecular dynamics of core excited states, as well as condensed phase simulations. ROKS can also model doubly excited/ionized states with one broken electron pair, which are beyond the ability of linear response based methods.
	\end{abstract}
	\maketitle

Spectroscopy of core electrons is an useful tool for characterizing local electronic structure in molecules and extended materials, and has consequently seen wide use for studying both static properties\cite{wilson2013x,hahner2001order,guo2011electronic} and dynamics\cite{chergui2017photoinduced,bhattacherjee2018ultrafast,schnorr2019tracing} of chemical systems. Theoretical modeling of core excited states is however a challenging task, as traditional quantum chemistry methods are typically geared towards understanding behavior of valence electrons. Indeed, it is common practice to `shift' computed X-ray absorption spectra (XAS) by several eV to align with experiment\cite{besley2010time,wenzel2014calculating,attar2017femtosecond,bhattacherjee2018photoinduced,hanson2019scaled,seidu2019simulation}. Such uncontrolled translation of spectra for empirical mitigation of systematic error is quite unappealing, and creates considerable scope for incorrect assignments. 

Linear response (LR) methods like time dependent density functional theory (TDDFT)\cite{runge1984density,casida1995time,dreuw2005single} and equation of motion coupled cluster (EOM-CC)\cite{stanton1993equation,krylov2008equation} are widely used to model excitations. LR methods do not require prior knowledge about the nature of targeted states, as they permit simultaneous calculation of multiple states on an even footing. However, widely used LR methods only contain a limited description of orbital relaxation, leading to poor performance for cases where such effects are essential (such as double excitations \cite{tozer2000determination,maitra2004double,loos2019reference}, as well as charge-transfer\cite{dreuw2005single,peach2008excitation} and Rydberg states\cite{casida1998molecular,tozer2000determination} in the case of TDDFT). Core excitations in particular are accompanied by substantial relaxation of the resulting core-hole (as well as relaxation of the valence density in response), leading to rather large errors with standard LR protocols. For instance, TDDFT spectra often need to be blue-shifted by $ > 10$ eV to correspond to experiment\cite{besley2010time,wenzel2014calculating,attar2017femtosecond,bhattacherjee2018photoinduced} (unless short-range corrected functionals specifically trained to predict core spectra are employed\cite{besley2009time}) and even EOM-CC singles and doubles (EOM-CCSD)\cite{stanton1993equation} tends to systematically overestimate energies by $1-2$ eV\cite{coriani2012coupled,frati2019coupled}. It worth noting that  second order algebraic diagrammatic construction (ADC(2)\cite{wormit2014investigating}, specifically CVS-ADC(2)-x\cite{wenzel2014calculating}) has been able to attain better accuracy for core-excitations, but only via compensation of basis set incompleteness errors with lack of orbital relaxation\cite{wenzel2014calculating}.  The O$(N^5)$ computational scaling of ADC(2) also restricts  applicability to large systems, relative to the lower scaling of DFT. \textcolor{black}{Methods based on the GW approximation\cite{hedin1965new} and the Bethe-Saltpeter equation (BSE)\cite{salpeter1951relativistic} have also been employed to study core spectra\cite{shirley2004ti,vinson2011bethe,gilmore2015efficient}}.

Orbital optimized (OO) methods attempt to incorporate the full effect of orbital relaxation on target excited states. The state specificity of OO methods necessitate prior knowledge about the nature of targeted states, making them not truly black-box. They have also historically been prone to `variational collapse' down to the ground state (as excited states are usually optimization saddle points), though recent advances in excited state orbital optimization have mitigated this to a great extent\cite{gilbert2008self,shea2018communication,ye2017sigma,hait2019excited}. OO methods have nonetheless been employed successfully for core ionizations\cite{zheng2019performance,lee2019excited,oakley2018deltadft,kahk2019accurate} and core excitations \cite{besley2009self,derricotte2015simulation,michelitsch2019efficient}.
There also exist LR methods that incorporate partial OO character, like Static Exchange (STEX)\cite{aagren1997direct} or Non-orthogonal Configuration Interaction Singles (NOCIS)\cite{oosterbaan2018non,oosterbaan2019non}, though such treatments are wave function based and $\sim 1$ eV error remains common due to lack of dynamic correlation. \textcolor{black}{Accurate single-point energies obtained from OO methods can also be employed to non-empirically translate LR excitation spectra into better agreement with experiment\cite{leetmaa2006recent,leetmaa2010theoretical}.}

The most widely used OO approach for modeling core excitations is $\Delta$ Self-Consistent Field ($\Delta$SCF)\cite{ziegler1977calculation,gilbert2008self,kowalczyk2011assessment,besley2009self}, where a non-aufbau solution to the Hartree-Fock\cite{szabo2012modern} or Kohn-Sham\cite{kohn1965self} DFT equations is converged. Unfortunately, single excitations in closed shell molecules cannot be represented by a single Slater determinant, resulting in spin-contaminated ``mixed" $\Delta$SCF solutions that are intermediate between singlet and triplet.  The core-hole and the excited electron are nonetheless expected to be fairly independent (due to low spatial overlap between orbitals), and spin-contaminated $\Delta$SCF solutions can therefore be reasonably purified to a singlet via approximate spin-projection (AP)\cite{yamaguchi1988spin}. AP however entails independent optimization of the triplet state, resulting in two sets of orbitals per targeted singlet state, which is both computationally inefficient and intellectually unappealing. Furthermore, spin-unrestricted DFT can exhibit rather unusual catastrophic failures with electronic configurations far from equilibrium\cite{hait2019wellbehaved}, making a restricted approach preferable. 

Restricted Open-Shell Kohn-Sham\cite{filatov1999spin,kowalczyk2013excitation}(ROKS) solves both of these issues via optimizing $2E_{M}-E_{T}$ for the same set of spin-restricted (RO) orbitals ($E_{M}$ is the energy of the mixed Slater determinant and $E_{T}$ is the energy of the corresponding triplet determinant within the $M_s=1$ manifold). Most ROKS implementations (such as the one described in Ref  \onlinecite{kowalczyk2013excitation}) however tend to collapse down to the lowest excited singlet ($S_1$) state, hindering use for studying core excitations. The recently developed Square Gradient Minimization (SGM) approach\cite{hait2019excited} permits ROKS to target arbitrary singlet excited states with one broken electron pair, thereby making application to core excitations feasible. SGM has been described in detail elsewhere\cite{hait2019excited}, and we only note that each SGM iteration has a cost that ranges between twice (for methods with analytical orbital Hessians for the energy/Lagrangian) and thrice (for methods without such Hessians, necessitating use of a finite-difference approximation) the cost of evaluating the orbital gradient of the energy/Lagrangian. ROKS calculations with SGM therefore have the same scaling as ground state DFT calculations with methods like GDM\cite{van2002geometric} or DIIS\cite{pulay1980convergence}, but with a slightly larger prefactor per iteration. \textcolor{black}{SGM is also robust against variational collapse and can converge to excited states where the more conventional Maximum Overlap Method (MOM)\cite{gilbert2008self} encounters variational collapse or fails to converge\cite{hait2019excited}.}

\begin{table}[h!]
\small{\begin{tabular}{lllllllll}
\hline
Species & Core orbital & Expt. & SPW92 & PBE   & B97M-V & SCAN  & PBE0  & $\omega$B97X-V \\ \hline
C$_2$H$_4$      & C      & 284.7\cite{hitchcock1977carbon} & 281.1 & 284.0 & 286.4 & 284.7 & 284.3 & 285.1 \\
HCHO      & C      & 285.6\cite{remmers1992high} & 282.1 & 284.9 & 287.4 & 285.7 & 285.2 & 286.0 \\
C$_2$H$_2$      & C      & 285.9\cite{hitchcock1977carbon} & 282.1 & 284.8 & 287.3 & 285.7 & 285.2 & 286.0 \\
C$_2$N$_2$      & C      & 286.3\cite{hitchcock1979inner} & 282.5 & 285.3 & 287.8 & 286.2 & 285.7 & 286.6 \\
HCN       & C      & 286.4\cite{hitchcock1979inner} & 282.8 & 285.5 & 288.0 & 286.3 & 285.8 & 286.6 \\
(CH$_3$)$_2$CO     & C (CO) & 286.4\cite{prince2003near} & 282.9 & 285.6 & 288.1 & 286.4 & 285.9 & 286.6 \\
C$_2$H$_6$      & C      & 286.9\cite{hitchcock1977carbon} & 282.8 & 285.8 & 288.1 & 286.7 & 286.3 & 287.3 \\
CO        & C      & 287.4\cite{domke1990carbon} & 283.5 & 286.1 & 288.7 & 287.0 & 286.5 & 287.3 \\
CH$_4$       & C      & 288.0\cite{schirmer1993k} & 284.0 & 286.9 & 289.4 & 288.0 & 287.4 & 288.5 \\
CH$_3$OH      & C      & 288.0\cite{prince2003near} & 284.6 & 287.5 & 289.9 & 288.2 & 287.7 & 288.7 \\
HCOOH     & C      & 288.1\cite{prince2003near} & 284.2 & 287.0 & 289.6 & 288.0 & 287.4 & 288.2 \\
HCOF      & C      & 288.2\cite{robin1988fluorination} & 284.4 & 287.2 & 289.8 & 288.1 & 287.6 & 288.4 \\
CO$_2$       & C      & 290.8\cite{prince1999vibrational} & 286.5 & 289.1 & 292.0 & 290.3 & 289.7 & 290.5 \\
CF$_2$O      & C      & 290.9\cite{robin1988fluorination} & 286.8 & 289.5 & 292.3 & 290.6 & 290.0 & 290.8 \\
C$_2$N$_2$      & N      & 398.9\cite{hitchcock1979inner} & 394.5 & 397.8 & 400.5 & 398.7 & 398.2 & 399.1 \\
HCN       & N      & 399.7\cite{hitchcock1979inner} & 395.4 & 398.7 & 401.3 & 399.5 & 399.0 & 399.8 \\
Imidazole & N (CH=N-CH)     & 399.9\cite{apen1993experimental} & 395.6 & 398.9 & 401.5 & 399.7 & 399.2 & 399.9 \\
NH$_3$       & N      & 400.8\cite{schirmer1993k} & 395.9 & 399.4 & 402.0 & 400.3 & 399.8 & 400.9 \\
N$_2$        & N      & 400.9\cite{myhre2018theoretical} & 396.6 & 399.8 & 402.5 & 400.7 & 400.1 & 400.9 \\
N$_2$O       & N ($\overset{*}{\mathrm{N}}$NO)     & 401.0\cite{prince1999vibrational} & 396.7 & 400.0 & 402.7 & 400.9 & 400.2 & 401.0 \\
Glycine   & N      & 401.2\cite{plekan2007x} & 396.5 & 400.0 & 402.6 & 400.9 & 400.5 & 401.6 \\
Pyrrole   & N      & 402.3\cite{pavlychev1995nitrogen} & 397.8 & 401.3 & 403.9 & 402.2 & 401.7 & 402.5 \\
Imidazole & N (CH-NH-CH)     & 402.3\cite{apen1993experimental} & 397.9 & 401.3 & 403.9 & 402.2 & 401.7 & 402.5 \\
N$_2$O       & N (N$\overset{*}{\mathrm{N}}$O)     & 404.6\cite{prince1999vibrational} & 400.0 & 403.3 & 406.1 & 404.4 & 403.7 & 404.5 \\
HCHO      & O      & 530.8\cite{remmers1992high} & 525.9 & 529.8 & 532.5 & 530.6 & 529.9 & 530.8 \\
(CH$_3$)$_2$CO     & O      & 531.4\cite{prince2003near} & 526.2 & 530.1 & 532.8 & 531.0 & 530.3 & 531.1 \\
HCOF      & O      & 532.1\cite{robin1988fluorination} & 527.0 & 530.9 & 533.6 & 531.8 & 531.0 & 531.9 \\
HCOOH     & O(CO)  & 532.2\cite{prince2003near} & 526.9 & 530.8 & 533.5 & 531.7 & 530.9 & 531.8 \\
CF$_2$O      & O      & 532.7\cite{robin1988fluorination} & 527.9 & 531.9 & 534.7 & 532.8 & 532.0 & 532.9 \\
H$_2$O       & O      & 534.0\cite{schirmer1993k}         & 528.6 & 532.5 & 535.4 & 533.6 & 533.0 & 534.0 \\
CH$_3$OH      & O      & 534.1\cite{prince2003near}       & 528.8 & 532.7 & 535.5 & 533.8 & 533.2 & 534.1 \\
CO        & O      & 534.2\cite{domke1990carbon} & 529.1 & 533.0 & 535.7 & 533.9 & 533.1 & 534.0 \\
N$_2$O       & O      & 534.6\cite{prince1999vibrational} & 529.9 & 533.9 & 536.7 & 534.8 & 533.9 & 534.6 \\
Furan     & O      & 535.2\cite{duflot2003core} & 530.0 & 534.0 & 536.6 & 534.9 & 534.2 & 535.1 \\
HCOOH     &  O(OH)     & 535.4\cite{prince2003near} & 530.1 & 534.2 & 537.0 & 535.2 & 534.5 & 535.4 \\
CO$_2$       & O      & 535.4\cite{prince1999vibrational} & 530.3 & 534.2 & 537.1 & 535.3 & 534.4 & 535.3 \\
F$_2$        & F      & 682.2\cite{hitchcock1981k} & 676.8 & 681.2 & 683.9 & 682.0 & 681.1 & 682.0 \\
HF        & F      & 687.4\cite{hitchcock1981k} & 681.4 & 685.8 & 688.9 & 687.1 & 686.2 & 687.1 \\
HCOF      & F      & 687.7\cite{robin1988fluorination} & 681.8 & 686.3 & 689.3 & 687.5 & 686.5 & 687.5 \\
CF$_2$O      & F      & 689.2\cite{robin1988fluorination} & 683.4 & 687.9 & 691.0 & 689.1 & 688.1 & 689.1  \\ \hline
\end{tabular}}
\caption{Comparison between experimental (Expt.) and ROKS/aug-cc-pCVTZ K edge  (lowest symmetry allowed transition from 1s core orbitals) vertical absorption energies, of 40 core excitations in small molecules (in eV). }
\label{tab:roksdata}
\end{table}

A rather important consideration for use of ROKS is the choice of a functional out of the vast DFT alphabet soup. This is especially relevant for core spectroscopy, as modern DFT functionals have been trained/assessed mostly on modeling ground state energetics\cite{mardirossian2018survival,goerigk2017look,najibi2018nonlocal,mardirossian2018survival} and properties\cite{hait2018accurate,hait2018accuratepolar,hait2018delocalization}, which only depend on behavior of \textit{valence} electrons. It therefore seems appropriate to consider non-empirical density functionals like LSDA\cite{PW92}, PBE\cite{PBE} and PBE0\cite{pbe0}, or minimally parametrized functionals like SCAN\cite{SCAN} or $\omega$B97X-V\cite{wb97xv} that are fairly strongly constrained within functional space. It also seems worthwhile to assess the performance of highly accurate modern functionals like B97M-V\cite{b97mv}, that are less tightly constrained. We have consequently examined the performance of these six functionals in predicting ROKS excitation energies for 40 K edge transitions (i.e. from the 1s orbital) of C,N,O and F, for which relativistic effects are expected to be small. The resulting values have been listed in in Table \ref{tab:roksdata}, while statistical measures of error have been provided in Table \ref{tab:errors}. Table \ref{tab:errors} also lists errors in core ionization potentials (core IPs) and term values (gap between K edge and core IP), in order to give a more complete idea about the full spectrum. \textcolor{black}{This indirect measure is useful, since it is often difficult to identify individual transitions beyond the edge from experimental spectra. We do however note that ROKS/SGM can converge to higher excited states beyond the K edge with ease, preserving similar levels of accuracy as the K edge predictions (examples provided in Supporting Information).}

\begin{table}[thb!]
\begin{tabular}{|l|rr|rr|rr|rr|}
\hline
Functional & \multicolumn{2}{c}{K edge} \hfill \vline & \multicolumn{2}{c}{K edge (+rel. corr.)} \hfill \vline& \multicolumn{2}{c}{Core IP (+rel. corr.)} \hfill \vline& \multicolumn{2}{c}{Term value}\hfill \vline \\ \hline
           & RMSE      & ME        & RMSE    & ME    & RMSE   & ME   & RMSE       & ME       \\
SPW92      & 4.6       & -4.6      & 4.4           & -4.3        & 4.2         & -4.2      & 0.3        & 0.2      \\
PBE        & 1.2       & -1.1      & 0.9           & -0.9        & 0.8         & -0.8      & 0.3        & 0.1      \\
B97M-V     & 1.6       & 1.6       & 1.8           & 1.8         & 1.8         & 1.8       & 0.3        & 0.1      \\
SCAN       & 0.2       & -0.2      & 0.2           & 0.1         & 0.3         & 0.2       & 0.3        & 0.2      \\
PBE0       & 0.9       & -0.8      & 0.6           & -0.6        & 0.4         & -0.4      & 0.4        & 0.2      \\
$\omega$B97X-V    & 0.2       & 0.1       & 0.4           & 0.3         & 0.5         & 0.4       & 0.4        & 0.1      \\ \hline
\end{tabular}
\caption{Root mean squared error (RMSE) and mean error (ME) for prediction of K edge energies listed in Table \ref{tab:roksdata} (in eV). The effect of relativistic corrections (rel. corr.) have also been considered. The errors in prediction of the corresponding core ionization potential (core IP) and the term value (difference between K edge and core IP) are also reported. }
\label{tab:errors}
\end{table}

The values in Table \ref{tab:errors} make it quite clear that the SCAN and $\omega$B97X-V functionals are highly accurate in predicting the K edge, having an RMSE on the order $0.3$ eV irrespective of the presence of atom specific relativistic shifts. $\omega$B97X-V appears to be a bit less accurate for the prediction of core IPs than SCAN, but the greater variation in experimental measurements of core IPs\cite{jolly1984core} indicates that not too much meaning should be drawn from this. The classic PBE0 functional also appears to be fairly accurate when relativistic effects are included (although the K edge RMSE is about twice as large as that of $\omega$B97X-V). The SPW92\cite{Slater,PW92} LSDA functional systematically underestimates energies by $>4$ eV, on account of it only being exact for the uniform electron gas and therefore incapable of modeling the inhomogeneities present in the densities of core excited states. The PBE generalized gradient approximation (GGA) systematically underestimates energies by about an eV, while the B97M-V meta-GGA surprisingly appears to systematically \textit{overestimate} by $>1.5$ eV. Finally, all functionals predict term values to approximately the same accuracy, indicating that empirically translating ROKS spectra by functional specific constant shifts would lead to similar levels of accuracy, irrespective of the functional used. We however feel that uncontrolled translation of spectra is rather unappealing and will not pursue that avenue further. 

The high accuracy predicted by SCAN and $\omega$B97X-V (relative to experimental errors, which are on the order of 0.1 eV) merits further analysis to determine the factors responsible, and what error cancellations (if any) are occurring. Some of the most obvious factors to consider are relativistic effects, the roles played by orbital relaxation and delocalization error, as well as basis set incompleteness errors. Scalar relativistic effects systematically bind core electrons tighter than what predictions from non-relativistic DFT should suggest. The magnitude of this correction can be estimated from the difference between core IPs calculated with relativistic and non-relativistic theories for bare atoms. This approximation should be fairly accurate for second period elements, as the chemical environment would only slightly perturb these already small corrections (the reported values \cite{takahashi2017relativistic} range from 0.1 eV for C to 0.6 eV for F). Inclusion of these relativistic shifts however has minimal impact on the K edge RMSE for SCAN and $\omega$B97X-V (as can be seen from Table \ref{tab:errors}), as well as for core IPs (as shown in the Supporting Information).  The corrections do however appear to perceptibly lower RMSE for PBE0, by reducing some of the systematic underestimation. We also note that relativistic corrections are expected to be much larger past the second period, and cannot be neglected in K edge studies of heavier atoms. 

\begin{table}[htb!]
\begin{tabular}{lll}
 \hline             & HF    & CH$_4$   \\\hline
Experiment    & 687.4 & 288.0 \\
SCAN/TDDFT    & 666.1 & 273.8 \\
SCAN/$\Delta$SCF     & 687.1 & 287.9 \\
SCAN/ROKS     & 687.0 & 288.0 \\
$\omega$B97X-V/TDDFT & 668.7 & 276.5 \\
$\omega$B97X-V/$\Delta$SCF  & 687.2 & 288.5 \\
$\omega$B97X-V/ROKS  & 687.1 & 288.5
\end{tabular}
\caption{Comparison of TDDFT, $\Delta$SCF and ROKS K edges (in eV) for HF and CH$_4$ with the SCAN/$\omega$B97X-V functionals and the aug-cc-pCVTZ basis. The $\Delta$SCF values have been spin-purified with AP.}
\label{tab:tddft}
\end{table}

The overall effect of explicit orbital optimization via ROKS can be gauged by comparison to LR-TDDFT. Table \ref{tab:tddft} presents the results for the CH$_4$ and HF molecules, which conclusively demonstrate the utility of orbital optimization (as TDDFT underestimates experiment by 15-20 eV). We also note that $\Delta$SCF has similar accuracy as ROKS, showing that the coupling between the core-hole and excited electron is indeed very weak. Our conclusions about the behavior of ROKS with various functionals are therefore likely transferable to $\Delta$SCF in the regimes where the latter does not exhibit any unphysical behavior. 

The poor performance of TDDFT naturally raises questions about the role of delocalization error\cite{perdew1982density} (of which self-interaction error is but one part\cite{mori2006many,hait2018delocalization}), which is the factor typically responsible for systematic underestimation of TDDFT excitation energies\cite{dreuw2005single}. The excellent behavior of the SCAN meta-GGA local functional, and the relatively small performance gap between the local PBE and the global hybrid PBE0 functionals seem to suggest that delocalization error is not a major factor for ROKS. This is consistent with earlier observations of ROKS predicting excellent charge-transfer\cite{hait2016prediction} and Rydberg\cite{hait2019excited} state energies for cases where TDDFT performs poorly. Delocalization error of course continues to exist for ROKS, but orbital optimization drastically reduces the magnitude of delocalization driven errors that LR methods tend to predict\cite{ziegler2008revised,subotnik2011communication,hait2016prediction}, down to ground state calculation levels. 

\begin{table}[htb!]
\begin{tabular}{llll}
\hline
           & Delocalized hole & Localized hole& Difference \\ \hline
SPW2       & 388.4       & 396.6     & -8.2                 \\
PBE        & 391.7       & 399.8     & -8.1                 \\
B97M-V     & 398.9       & 402.5     & -3.6                 \\
SCAN       & 395.3       & 400.7     & -5.4                 \\
PBE0       & 396.4       & 400.1     & -3.7                 \\
$\omega$B97X-V    & 395.7       & 400.9     & -5.3                 \\
Experiment &             & 401.0     &                      \\ \hline
\end{tabular}
\caption{Comparison of the N$_2$ K edge predicted by ROKS/aug-cc-pCVTZ (in eV) between the fully delocalized and fully localized core-hole limits.}
\label{tab:deloc}
\end{table}

There is however an additional subtlety associated with systems possessing chemically identical atoms (like N$_2$ or O in CO$_2$), where the core-hole densities \textcolor{black}{of exact eigenstates} should be delocalized over multiple sites on account of symmetry. The coupling between core orbitals is nonetheless quite weak and localized core-hole diabatic states are therefore expected to be energetically quite close (i.e. within order of 0.01 eV)\cite{oosterban2019onecenter} to symmetric eigenstates. The energies of delocalized states in DFT are typically systematically underestimated on account of delocalization error (even within an OO framework), making use of localized core-hole states preferable for calculating core excitation energies. A quantitative measure of this effect for the N$_2$ molecule has been supplied in Table \ref{tab:deloc}. In practice therefore, the spurious delocalization effect should be avoided by supplying a localized core-hole as the initial guess and letting SGM converge to the closest localized solution. \textit{However}, it means that canonical orbitals cannot be used as initial guesses due to their inherently delocalized nature, and some localization scheme (or even a weak, symmetry breaking electric field) must be employed to generate initial guess orbitals for ROKS. It is somewhat intellectually unsatisfying to completely neglect delocalized states (which appear to be the lowest energy ROKS core-hole states \textcolor{black}{as well as representative of the behavior expected from true eigenstates}), but this pragmatic choice is essential in light of known failures of DFT for delocalized states\cite{Dutoi2006,Ruzsinszky2006,hait2018accurate,hait2018communication}. Fully symmetric states can be obtained from a NOCI approach\cite{oosterbaan2018non,oosterbaan2019non,oosterban2019onecenter}, but such multireference techniques cannot be straightforwardly generalized to DFT. We additionally note that localized orbitals has long been employed to improve the performance of wave function based approaches as well\cite{oosterbaan2018non,cederbaum1986double}, although use of delocalized orbitals therein lead to \textit{higher} energies (on account of missing correlation\cite{hait2018delocalization}). \textcolor{black}{The actual energy gap between the exact eigenstate with a delocalized core-hole and a localized core-hole state is however quite small overall\cite{oosterban2019onecenter}, and therefore use of localized ROKS solutions is an acceptable pragmatic choice. We additionally note that this small gap indicates that any experimental realization of a localized core-hole state (due to finite-temperature effects or other symmetry breaking) in experiment would not affect accuracy of experimental data employed in this study.}

\begin{table}[htb!]
\begin{tabular}{lllllll}
\hline
         & \multicolumn{2}{l}{aug-cc-pCVDZ} & \multicolumn{2}{l}{aug-cc-pCVTZ} & \multicolumn{2}{l}{aug-cc-pCVQZ} \\ \hline
         & Core IP         & K edge         & Core IP         & K edge         & Core IP         & K edge         \\
CH$_4$ (C)  & 292.07&	289.42&	291.16&	288.50&	291.11&	288.44\\
NH$_3$ (N)  & 407.02          & 402.01         & 405.83          & 400.85         & 405.76          & 400.78         \\
H$_2$O (O)  & 541.33          & 535.44         & 539.86          & 533.99         & 539.75          & 533.88         \\
HF (F)   & 695.63          & 688.86         & 693.87          & 687.13         & 693.72          & 686.98         \\
HCHO (C) & 295.89          & 286.97         & 294.97          & 286.03         & 294.91          & 285.97         \\
HCHO (O) & 540.76          & 532.32         & 539.27          & 530.83         & 539.15          & 530.71         \\ 
HCN (C) & 295.03 & 287.71         & 293.9 & 286.58         &  293.84 & 286.51  \\
HCN (N) &     408.37 & 401.09     & 407.07 & 399.8         & 406.99 & 399.71         \\ \hline
\end{tabular}
\caption{Convergence of $\omega$B97X-V core ionization potential (IP) and K edge absorption energies (in eV) against basis set size.}
\label{tab:corerelax}
\end{table}

The final factor we consider is basis set incompleteness error, whose analysis would also assist basis set selection for realistically sized systems (as aug-cc-pCVTZ is too impractically large). Valence excitation energies typically do not exhibit very strong basis set dependence\cite{loos2018mountaineering}, but the situation for core spectra is different due to the need to adequately relax the core-hole. Table \ref{tab:corerelax} compares the $\omega$B97X-V core ionization and K edge energies with increasing basis set cardinality. The small difference between aug-cc-pCVTZ and aug-cc-pCVQZ and the exponential convergence of SCF energies\cite{jensen2005estimating} suggest that aug-cc-pCVQZ values are functionally at the complete basis set limit. It can also be seen that aug-cc-pCVTZ systematically overestimates energies by about 0.1 eV relative to aug-cc-pCVQZ. This deviation is non-negligible relative to the low RMSE of SCAN and $\omega$B97X-V, but is quite comparable to the error bars inherent in experiment, indicating that the basis set incompleteness error in Table \ref{tab:roksdata} is not particularly significant. We nonetheless note that the slight overestimation of energies by aug-cc-pCVTZ seems to suggest that a component of the systematic overestimation of energies (after relativistic corrections) for SCAN and $\omega$B97X-V stems from basis set truncation, suggesting slightly lower errors at the complete basis set limit. 

Table \ref{tab:corerelax} also makes it apparent that aug-cc-pCVDZ is too small for accurate predictions, as energies are systematically overestimated  by 1-2 eV. The core IP is overestimated by almost the same amount as the K edge, indicating that the basis set incompleteness effects essentially arise from insufficient core relaxation alone. We therefore recommend that a mixed basis strategy be employed for larger species (where full aug-cc-pCVTZ is impractical), wherein the localized target atom employs a split core-valence triple zeta (CVTZ) basis, while the remaining atoms are treated with some smaller basis. A similar mixed basis approach has also been reported in literature \cite{ambroise2018probing}. This strategy (using aug-cc-pCVTZ in the target site and aug-cc-pVDZ for other atoms) reproduced the full $\omega$B97X-V/aug-cc-pCVTZ results for both the C and O K edges of HCHO to $\le 0.02$ eV deviation, suggesting its general efficacy \textcolor{black}{for predicting K edges of second period elements}. \textcolor{black}{It is however important to recognize that CVnZ bases are not likely to be sufficiently flexible in describing 1s electrons beyond the second period, and more specialized (or even uncontracted) basis sets may prove necessary for the local site of the K shell excitation for heavy elements.} In addition, we note that while diffuse functions are not strictly necessary for excitations to antibonding orbitals, they are critical for Rydberg states, with double augmentation being necessary to converge the higher core excited states of small molecules (as shown in the Supporting Information).

\begin{figure}[thb!]
    \centering
    \includegraphics[width=0.6\textwidth]{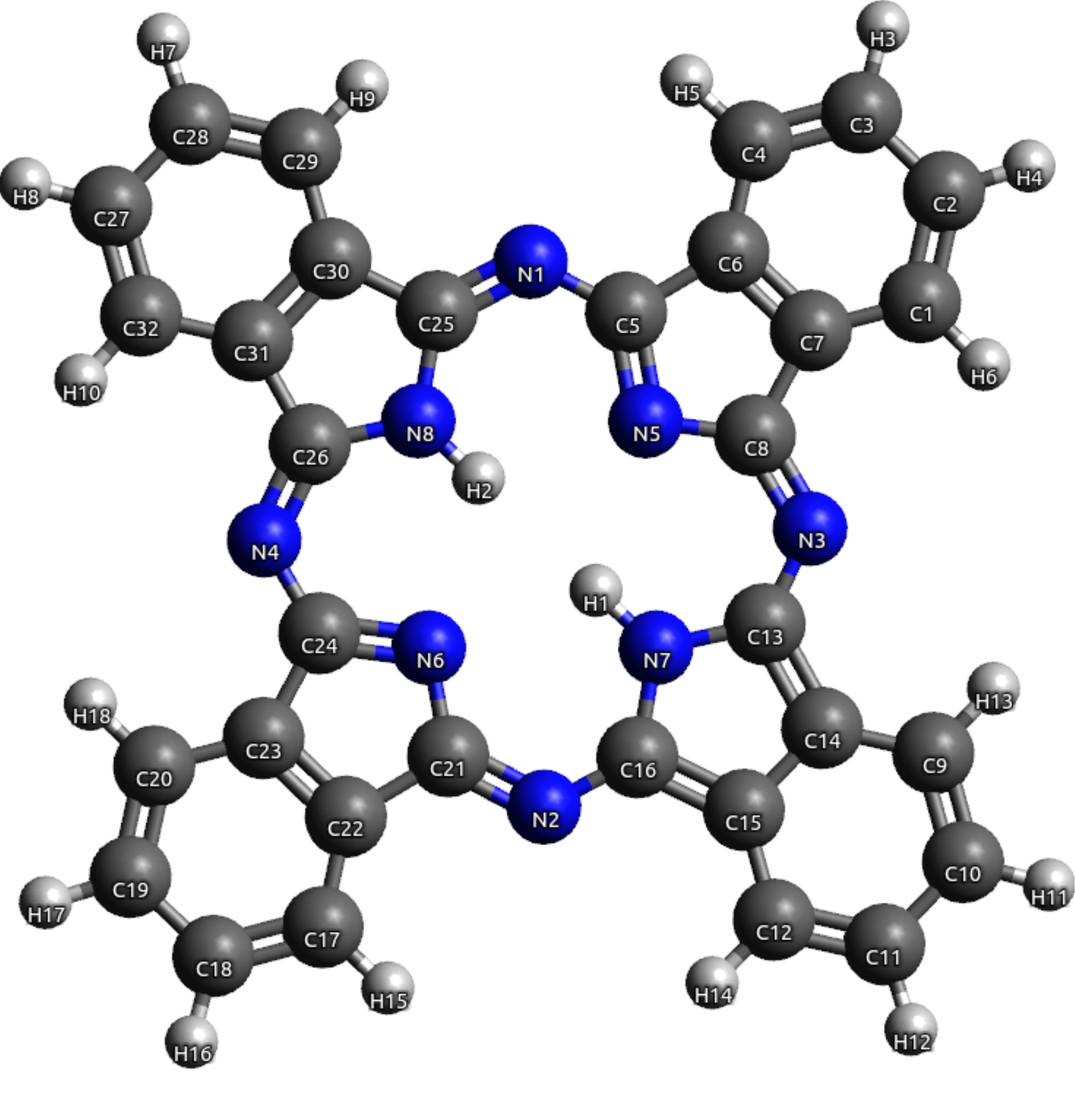}
    \caption{H$_2$Pc molecule.}
    \label{fig:h2pc}
\end{figure}

We next demonstrate the viability of applying ROKS/SGM to sizeable systems by computing the N K edge of the phthalocyanine molecule (H$_2$Pc, depicted in Fig \ref{fig:h2pc}). We employ the mixed basis strategy described and validated earlier, with the large cc-pCVTZ basis being applied to the target site while all other atoms use cc-pVDZ. We note that an additional advantage of the mixed basis approach is that it automatically breaks chemical equivalence of the target site, thereby spontaneously localizing the resulting core orbital (sans explicit localization). Fig \ref{fig:h2pc} shows that H$_2$Pc has three different types of N atoms. N1-N4 are bridging aza type, N7-N8 are NH pyrrole like while N5-N6 are hydrogen free pyrrole like.  A comparison between $\omega$B97X-V/ROKS excitation energies and experimental values from thin film measurements\cite{kera2006high} are supplied in Table \ref{tab:h2pc}. We continue to find remarkably good agreement between theory and experiment, with the N core energies being predicted to be in the order N5$<$N1$<$N7. This is consistent with behavior observed for imidazole in Table \ref{tab:roksdata}.

\begin{table}[htb!]
\begin{tabular}{lll}
\hline
ROKS & core-hole site & Experiment  \\ \hline
398.3  & N5        & 397.9 \\
398.4  & N5        & 398.3 \\
398.5  & N1        &       \\
399.1  & N1        & 399   \\
400.3  & N7        & 399.7 \\
400.5  & N7        & 400.3 \\ \hline
\end{tabular}
\caption{Comparison of experimental N 1s excitation energies\cite{kera2006high} (in eV) of H$_2$Pc to predictions from ROKS with $\omega$B97X-V. A mixed basis set (see text) was used.}
\label{tab:h2pc}
\end{table}

\begin{table}[htb!]
\begin{tabular}{|l|llll|lll|}
\hline
     & \multicolumn{3}{c}{L$_3$}  &  & \multicolumn{3}{c}{L$_2$}  \\ \hline
     & Expt. & SCAN  & $\omega$B97X-V &  & Expt. & SCAN  & $\omega$B97X-V \\\hline
SiH$_4$     & 102.6\cite{hayes1972absorption} & 103.0 & 102.9 &  & 103.2\cite{hayes1972absorption} & 103.6 & 103.5 \\
Si(CH$_3$)$_4$    & 102.9\cite{bozek1987high} & 102.8 & 102.8 &  & 103.5\cite{bozek1987high} & 103.4 & 103.4 \\
SiF$_4$     & 106.1\cite{friedrich1980overlapping} & 106.2 & 106.1 &  & 106.7\cite{friedrich1980overlapping} & 106.8 & 106.7 \\
$\overset{*}{\mathrm{Si}}$(Cl)$_4$  & 104.2\cite{bozek1987high} & 104.5&	104.6&  & 104.8\cite{bozek1987high} & 105.1&	105.2 \\
Si(OCH$_3$)$_4$    & 104.8\cite{sutherland1993si}&	104.9	&105.1&&105.4\cite{sutherland1993si}&	105.5	&105.7 \\
PH$_3$      & 131.9\cite{liu1990high} & 132.1 & 131.8 &  & 132.8\cite{liu1990high} & 132.9 & 132.6 \\
PF$_3$      & 134.9\cite{neville1998inner} & 134.9 & 134.7 &  & 135.6\cite{neville1998inner} & 135.7 & 135.5 \\
P(CH$_3$)$_3$     & 132.3\cite{liu1990high} & 132.5 & 132.2 &  & 133.1\cite{liu1990high} & 133.3 & 133.0 \\
PF$_5$      & 138.2\cite{hu2007high} & 138.0 & 138.0 &  & 139.0\cite{hu2007high} & 138.8 & 138.8 \\
OPF$_3$     & 137.1\cite{neville1998inner} & 137.0 & 136.9 &  & 137.8\cite{neville1998inner} & 137.8 & 137.7 \\
H$_2$S      & 164.4\cite{guillemin2005fragmentation} & 164.7 & 164.3 &  & 165.6\cite{guillemin2005fragmentation} & 165.9 & 165.5 \\
SF$_6$      & 172.3\cite{hudson1993high} & 172.0 & 171.9 &  & 173.4\cite{hudson1993high} & 173.2 & 173.1 \\
(CH$_3$S)$_2$ & 164.1\cite{schnorr2019tracing} & 164.0 & 163.6 &  & 165.4\cite{schnorr2019tracing} & 165.2 & 164.8 \\
CS$_2$      & 163.3\cite{hedin2009x} & 163.4 & 162.5 &  & 164.4\cite{hedin2009x} & 164.6 & 163.7 \\
CSO      & 164.3\cite{ankerhold1997ionization} & 164.4 & 163.7 &  & 165.5\cite{ankerhold1997ionization} & 165.7 & 164.9 \\
HCl      & 200.9\cite{lablanquie2011evidence} & 201.0 & 200.5 &  & 202.4\cite{aksela1990decay} & 202.6 & 202.1 \\
Cl$_2$      & 198.2\cite{nayandin2001angle} & 198.2 & 197.7 &  & 199.8\cite{nayandin2001angle} & 199.9 & 199.3 \\
ClF$_3$     & 201.8\cite{sze1989inner} & 201.7 & 201.3 &  & 203.2\cite{sze1989inner} & 203.3 & 203.0 \\
CCl$_4$     & 200.3\cite{lu2008state} & 200.1 & 199.7 &  & 201.9\cite{lu2008state} & 201.7 & 201.3 \\
C$_6$H$_5$Cl     & 201.5\cite{hitchcock1978inner}&	201.4&	201.0 & &203.2\cite{hitchcock1978inner}&	203.1&	202.6\\\hline
RMSE     &        & 0.2   & 0.4   &  &        & 0.2   & 0.4   \\
ME       &        & 0.1   & -0.2
&  &        & 0.1   & -0.2 \\\hline
\end{tabular}
\caption{Comparison between experimental (Expt.) and ROKS L$_{2,3}$ edges  (lowest symmetry allowed transition from 2p core orbitals) vertical absorption energies (in eV). SiH$_4$, PH$_3$, H$_2$S and HCl employ aug-cc-pCVTZ, while the the mixed basis strategy described above was used for the remaining species (aug-cc-pCVTZ as the large local basis and aug-cc-pVDZ for other atoms). Scalar relatvistic corrections for these atoms are $<0.1$ eV\cite{takahashi2017relativistic} and were thus neglected. The protocol for incorporating spin-orbit coupling is described in the appendix.}
\label{tab:ledge}
\end{table}

\textcolor{black}{
Having discussed the applicability of using ROKS/SGM for the computation of 1s excitation energies for second period elements, we briefly consider the behavior of inner shell excitations for heavier atoms. Excitations out of the 2p orbitals are of particular interest for third period elements. The degeneracy of the 2p orbitals is however broken by spin-orbit coupling (which is larger in magnitude than any splitting introduced by molecular symmetry on core levels), which results in two peaks with intensities roughly in a 2:1 ratio, that correspond to excitations out of the $2p_{3/2}$ and $2p_{1/2}$ levels respectively. These peaks are called L$_3$ and L$_2$ respectively (in contrast to the higher energy L$_1$ peaks stemming from excitations out of the 2s level). This spin-orbit splitting cannot be reproduced by any non-relativistic theory like Kohn-Sham DFT. Like the scalar relativistic shifts employed earlier however, they are not sensitive to the chemical environment of a given atom. The spin-orbit effects of the electron excited to a higher energy orbital is also typically negligible on account of greater distance from the nucleus.  It is therefore possible to estimate the L$_3$-L$_2$ splitting for a specific atom either via relativistic wave function theories or experiment, and transfer those values for other species via use of (near-)degenerate perturbation theory, in conjunction with the non-relativistic values computed with ROKS (as described in the appendix). Table \ref{tab:ledge} supplies a comparison between values obtained with this method (employing the hereto best performing SCAN and $\omega$B97X-V functionals) with experiment for a few species. Both functionals appear to retain the level of accuracy observed for the second period K edge. It does however appear that $\omega$B97X-V performs a little worse than SCAN due to systematic underestimation of excitation energies. Nonetheless, it is apparent that this approach is quite promising for computing core spectra of 2p electrons in heavier elements, in addition to the second period K shell spectroscopy discussed so far.}

\begin{figure}[htb!]
    \centering
    \includegraphics[width=0.5\textwidth]{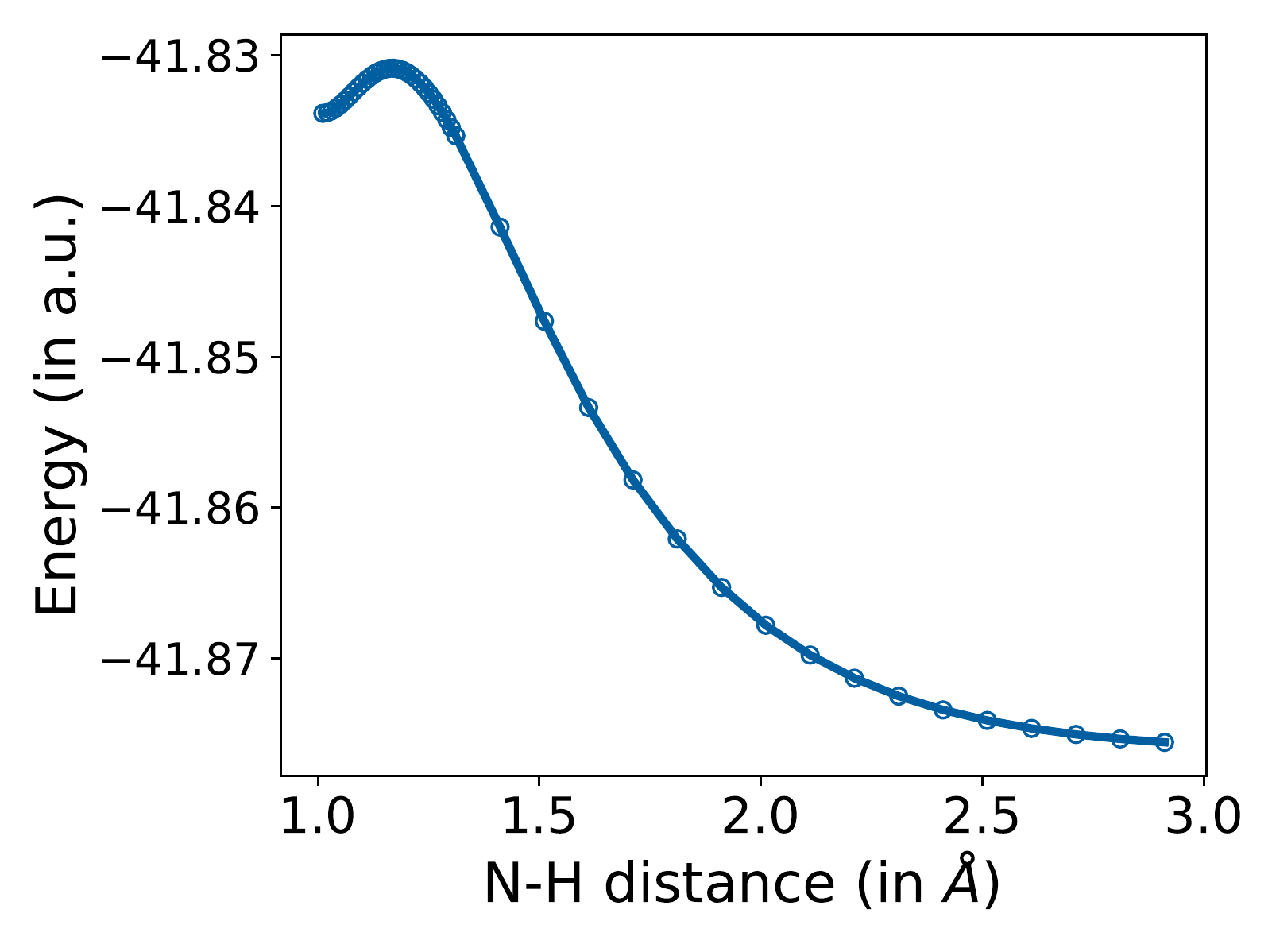}
    \caption{PES of core excited NH$_3$ (from ROKS $\omega$B97X-V/aug-cc-pCVTZ), against stretch of a NH bond. Nuclear positions of the other atoms were optimized for all points.}
    \label{fig:pes}
\end{figure}

It is also worth noting the analytical nuclear gradients for ROKS are fairly simple to obtain\cite{kowalczyk2013excitation}, permitting geometry optimizations and ab-initio molecular dynamics in the core-excited state (which could assist in studying ultrafast dissociation processes or lead to ab-initio computation of spectral linewidths, for instance). Conseuqently, it is also possible to compute vibrational spectra of core excited states via finite differences, making it possible to assign modes to vibrational fine structure of XAS. All of this can be acheived for the same computational scaling as ground state DFT, permitting application to very large systems. As a simple of proof of principle, Fig \ref{fig:pes} presents the potential energy surface (PES) of core excited NH$_3$ (1s$\to$ 4$a_1$) against NH stretching. This state can relax to a shallow local minimum, but ultrafast dissociation to NH$_2$+H is energetically more favorable (after crossing a small barrier\cite{walsh2015molecular}). ROKS is able to reproduce this behavior, which is a significant advantage over TDDFT (as the latter is completely incapable of modeling excited state bond dissociation\cite{hait2019beyond}). The barrier against dissociation is estimated to be 0.08 eV, which is within the $0.1$ eV error bar associated with the experimental estimate of 0.2 eV\cite{walsh2015molecular}. It is however worth noting that typical DFT error for ground state barrier prediction is of the order of 0.05 eV\cite{mardirossian2018survival}, and so ultraprecise predictions should not be realistically expected. The main strength of ROKS lies in that it can be applied to large systems with \textit{reasonable} accuracy. 

\begin{table}[htb!]
\small{\begin{tabular}{llllll}
\hline
                                        &  & CC3\cite{myhre2018theoretical} &SCAN& $\omega$B97X-V & Ground state \\ 
 & Expt. & d-aug-cc-pCVQZ & aug-cc-pCVTZ & aug-cc-pCVTZ & Expt.\\ \hline
Absorption energy (in eV)      & 400.90\cite{myhre2018theoretical}     & 401.03             & 400.73 & 400.91            & \\
Bond length (in {\AA}) & 1.164\cite{chen1989k}      & 1.158     & 1.154         & 1.147          &   1.098 \\
Frequency (in cm$^{-1}$)    & 1895\cite{chen1989k}  & 2032    & 2049      & 2134  & 2330	            \\ \hline
\end{tabular}}
\caption{Comparison of CC3 and ROKS predictions for first core excited state (1s$\to\pi^*$) state of N$_2$.}
\label{tab:vibdata}
\end{table}

We also demonstrate reasonable reproduction of the core excited state bond length and vibrational frequency of N$_2$ by ROKS, which has been fairly well characterized by both theory and experiment\cite{myhre2018theoretical}. A comparison with the experimental values, the CC3\cite{christiansen1995response} wave function method (from Ref \onlinecite{myhre2018theoretical}) and  ROKS is provided in Table \ref{tab:vibdata}. We find that theoretical methods predict a shorter and stiffer bond in the core-excited state, relative to experimental fits.  We do however note that the experimental values are not particularly precise, with the vibrational frequency being estimated from an experiment with a photon resolution of approx 50 meV (i.e. 403 cm$^{-1}$) and the bond length being calculated via a fit to a Morse potential\cite{chen1989k}, which does not appear to be consistent with coupled cluster studies\cite{myhre2018theoretical}. The SCAN predictions are in very good agreement with CC3, while $\omega$B97X-V predicts a shorter bond and higher vibrational frequency. This superficially seems to suggest higher reliability of SCAN geometries/frequencies, but considerable further testing is required before more general conclusions can be reached. At any rate, the low computational cost of ROKS with either functional makes it attractive relative to O($N^7$) scaling methods like CC3.  

\begin{table}[htb!]
\begin{tabular}{lllllllll}
\hline
Molecule & Hole site & Expt.   & SPW92 & PBE   & B97M-V & SCAN  & PBE0  & $\omega$B97X-V \\ \hline
C$_2$H$_2$  & C,C  & 596.0$\pm$ 0.5\cite{nakano2013single} & 588.3 & 593.8 & 598.7  & 595.6 & 594.7 & 596.3   \\
C$_2$H$_4$  & C,C  & 593.3$\pm$ 0.5\cite{nakano2013single} & 585.0 & 590.7 & 595.6  & 592.5 & 591.5 & 593.1   \\
C$_2$H$_6$  & C,C  & 590.0$\pm$ 0.5\cite{nakano2013single} & 581.7 & 587.6 & 592.4  & 589.3 & 588.3 & 589.9   \\
CO       & C,O  & 855.4$\pm$1\cite{nakano2013single} & 846.2 & 852.6 & 858.0  & 854.8 & 853.6 & 855.2   \\
CO$_2$      & C,O  & 848.6$\pm$ 1.2\cite{salen2012experimental} & 842.1 & 848.6 & 854.3  & 851.1 & 850.0 & 851.6   \\
N$_2$       & N,N  & 835.9$\pm$ 1\cite{nakano2013single}  & 827.9 & 834.4 & 839.9  & 836.7 & 835.7 & 837.3   \\
N$_2$O      & N,N  & 834.2$\pm$2.1\cite{salen2012experimental}  & 825.1 & 831.6 & 837.4  & 834.1 & 833.4 & 835.2   \\ \hline
\end{tabular}
\caption{Comparison between experimental and ROKS/aug-cc-pCVTZ TSDCH core ionization energies (in eV).}
\label{tab:tsdch}
\end{table}

It is also important to note that the ROKS is applicable to any singlet state with one broken electron pair\cite{hait2019excited}, and not just the single excitations considered so far. There is unfortunately very little high quality experimental data about doubly excited core states involving second period elements. We consequently look at two site double core-hole (TSDCH) states instead, which are intrinsically open-shell (possessing one unpaired electron in each singly ionized atomic site) and are thereby ideal candidates for ROKS. TSDCH states have been long proposed as sensitive measures of chemical environment\cite{cederbaum1986double}, leading to experimental effort towards their realization\cite{lablanquie2011evidence,salen2012experimental,nakano2013single}. We present a comparison between experimental and ROKS TSDCH ionization energies in Table \ref{tab:tsdch}. Similar behavior to the K edge data in Table \ref{tab:roksdata} is observed, with B97M-V massively overestimating, while SPW92/PBE underestimate. The large experimental error bars make it difficult to judge the relative performances of PBE0, SCAN and $\omega$B97X-V (however, $E_\textrm{PBE0} < E_\textrm{SCAN} < E_{\omega \textrm{B97X-V}}$ for all species). The predictions from the latter three functionals are overall quite reliable (considering the experimental error bars), and offer an inexpensive and spin pure way to compute TSDCH excitation energies (vs, say more expensive methods like $\Delta$CCSD(T), which does not lead to substantially enhanced accuracy for such systems\cite{lee2019excited}).  This certainly represents a major advantage of ROKS over TDDFT, which is incapable of modelling doubly excited states at all\cite{maitra2004double,levine2006conical,dreuw2005single}.

\begin{figure}[thb!]
\begin{minipage}{0.48\textwidth}
    \centering
    \includegraphics[width=\linewidth]{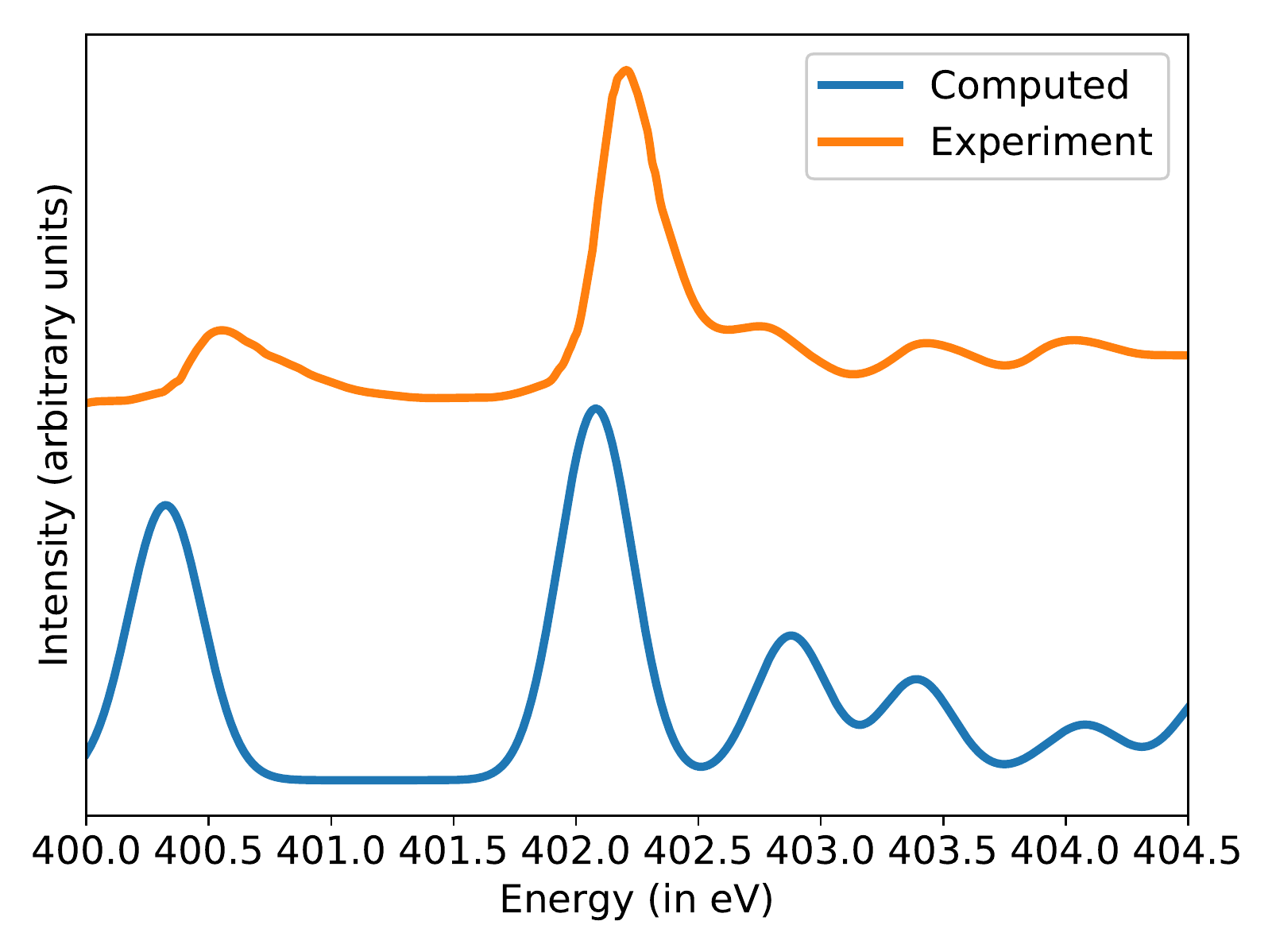}
    \subcaption{N in NH$_3$.}
\end{minipage}
\begin{minipage}{0.48\textwidth}
    \centering
    \includegraphics[width=\linewidth]{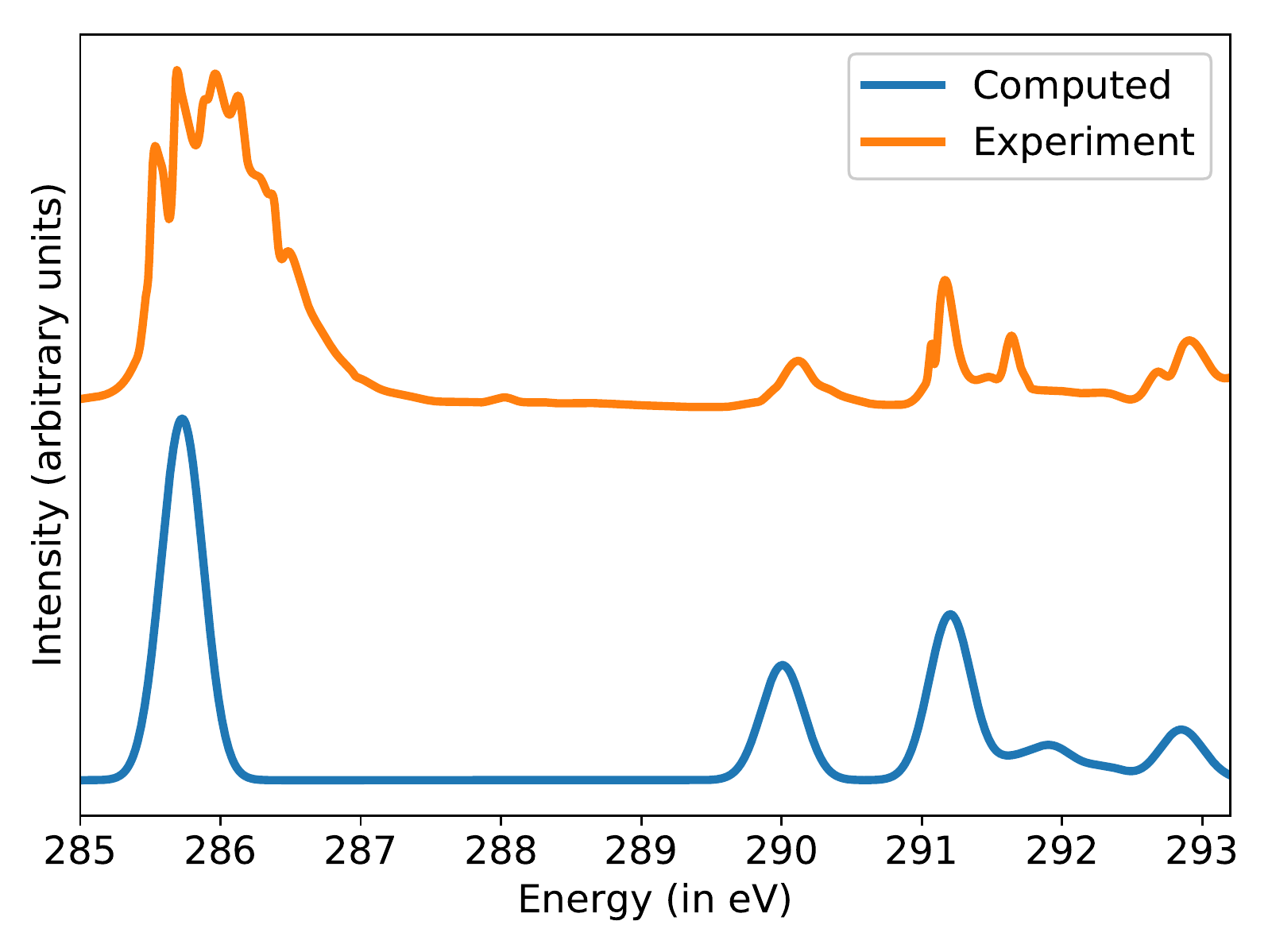}
    \subcaption{C in HCHO.}
\end{minipage}
    \caption{Spectra predicted by SCAN/d-aug-cc-pCVTZ compared to experiment (without any translational shift). Experimental data taken from Ref \citenum{schirmer1993k} for NH$_3$ and Ref \citenum{remmers1992high} for HCHO. Gaussian broadening was applied to the peaks, with $\sigma=0.15$ eV.}
    \label{fig:spectra}
\end{figure}

\textcolor{black}{Having described the accuracy of predicting energies via ROKS/SGM, we next briefly consider the prediction of full core absorption spectra. This is somewhat more of a challenge for OO based DFT methods, as transition properties like oscillator strengths cannot formally be defined within this framework (due to the fictitious nature of the Kohn-Sham determinant). Nonetheless, reasonable values can be obtained by approximating the wave function with the Kohn-Sham determinant, followed by computation of transition properties via a wave function like approach\cite{besley2009self}. While the actual values need not be very accurate (or obey formal properties like the Thomas-Reich-Kuhn rule\cite{dreuw2005single}), their relative variation is typically expected to be similar to exact behavior, resulting in roughly accurate spectral shape. Fig \ref{fig:spectra} presents the core excited spectra of N in NH$_3$ and C in HCHO against experimental results. The agreement is in no way perfect (on account of lack of nuclear quantum effects in the computed spectra, as well as use of uniform Gaussian broadening), but the main features are reproduced quite well and no translation of spectra is necessary at all. In particular, peaks corresponding to higher energy Rydberg states are quite visible, which clearly highlights SGM's ability to predict such states without variational collapse.} 

In summary, we demonstrate that single core excitation energies for the K shell of second period elements \textcolor{black}{and L$_{2,3}$ shells of third period elements} can be computed to $<0.5$ eV RMS error via the use of a state specific restricted open-shell Kohn-Sham (ROKS) approach, without any need to translate spectra at all. The computational scaling of ROKS is identical to the corresponding ground state DFT calculation (with a slightly larger prefactor), when it is combined with the recently developed square gradient minimization (SGM\cite{hait2019excited}) orbital optimizer, readily permitting application to large systems. The low ROKS errors owe greatly to advances in ground state density functional development, as modern functionals like SCAN and $\omega$B97X-V are found to be the most accurate. We further show that the low errors in prediction do not stem from basis set incompleteness errors or neglect of relativistic effects, indicating that ROKS is obtaining the right answer for the right reasons (namely that the excitation from one localized core orbital to the virtual space can be very well described by one configuration plus a description of dynamical correlation). The ready availability of analytic ROKS nuclear gradients also suggest considerable potential for employing this approach for excited state geometry optimization or ab-initio molecular dynamics. This is aided by the ability of ROKS to correctly describe excited state bond dissociations, unlike TDDFT. Finally, ROKS can be employed to double excitation or double ionization processes (where a total of one electron pair has been broken), which is difficult for LR methods. 

The high accuracy and low computational scaling of ROKS makes it an ideal method for studying the dynamics of both core excited states and XAS of valence excited states in sizeable systems. ROKS (with the local SCAN functional) is also an ideal method for simulating core spectra in the condensed phase. There does however exist a need to incorporate scalar relativistic effects, in order to extend applicability to \textcolor{black} {the innermost shells of heavy elements (where an additive atom specific correction might not be sensitive enough)}. Work along these directions is presently in progress.

\section*{Computational Details}
All calculations were performed with the Q-Chem 5.2 \cite{QCHEM4} package. 	Local exchange-correlation integrals were calculated over
a radial grid with 99 points and an angular Lebedev grid with 590 points. Core IPs were computed with RO-$\Delta$SCF, which is spin-pure and equivalent to ROKS when an electron is excited to infinity. The core-ionized RO-$\Delta$SCF orbitals were subsequently used as initial guesses for ROKS absorption energy calculations. This reduces number of ROKS iterations, by effectively decoupling the core-hole relaxation from the rest of the optimization. Such a strategy would be extremely useful for computing multiple excited states, as the core-hole relaxation process would need to be converged only once to generate initial guesses for several ROKS calculations. SGM was employed for all $\Delta$SCF/ROKS computations. Experimental geometries (from the NIST database\cite{johnson2015nist}) were used whenever possible, with geometries being optimized with MP2/cc-pVTZ in their absence (except for H$_2$Pc, where $\omega$B97X-V/def2-SV(P) was used instead). Vibrational frequencies $\omega$ in Table \ref{tab:vibdata} were found by solving the nuclear wave equation for the PES, and subsequent fitting to the anharmonic oscillator energy $E_{\nu}=\hbar \omega\left(\nu+\dfrac{1}{2}\right)-\hbar \omega x_e\left(\nu+\dfrac{1}{2}\right)^2$ 
\section*{Acknowledgment} 
	This research was supported by the Director, Office of Science, Office of Basic Energy Sciences, of the U.S. Department of Energy under Contract No. DE-AC02-05CH11231. D.H. would like to thank Scott Garner, Katherine Oosterbaan and Andrew Ross for stimulating discussions. 
\section*{Supporting Information} 
	\noindent Geometries of species studied (zip), Raw data (xlsx).
	
\appendix
\section*{Spin-orbit effects in L-edge spectra}
There are six 2p spin orbitals ($\ket{p_{x,y,z}}\otimes\ket{\uparrow,\downarrow}$ that are degenerate in non-relativistic quantum mechanics (for symmetric molecular fields and in the absence of magnetic fields). The spin-orbit coupling operator $-J\vec{L}\cdot\vec{S}$ breaks this degeneracy. It can be easily shown that the $\ket{p_z}\otimes\ket{\uparrow}$ couples with the $\ket{p_{x,y}}\otimes\ket{\downarrow}$ (and the reverse). Within this reduced subspace of 3 interacting orbitals, the spin-orbit coupling operator can be represented as:
\begin{align}
    \hat{H}^{(1)}&=-J \begin{pmatrix}
    0 & 1 & i\\
    1 & 0 & -i\\
    -i & i &0\\
    \end{pmatrix}
\end{align}
In most molecules however, the degeneracy between the spatial $p$ levels is broken due to (lack of) symmetry by a small (0.1 eV scale) amount. This effect should also ideally be accounted for, and so we have a full Hamiltonian:
\begin{align}
    \hat{H}&=\begin{pmatrix}
    \omega_1 & -J & -iJ\\
    -J & \omega_2 & iJ\\
    iJ & -iJ &\omega_3\\
    \end{pmatrix}
\end{align}
where $\omega_{1,2,3}$ are the non-relativistic excitation energies out of the three p orbitals (as computed with ROKS or some other method). Diagonalization of this $\hat{H}$ would yield the predicted $L$ excitation energies. 

The case of $\omega_{1,2,3}$ being degenerate (i.e. symmetric molecular field) yields the well known case where the eigenvalues are $\omega-J,\omega-J$ (L$_3$ band) and $\omega+2J$ (L$_2$ band). The 2:1 degeneracy ratio explains the standard 2:1 heights seen in experimental spectra. The separation between these (3$J$) is called the doublet splitting and is experimentally\cite{barrie1974correlation} found to be 0.6 eV for Si, 0.8 eV for P, 1.2 eV for S and 1.6 eV for Cl. A weak molecular field of the order of $0.1$ eV therefore is unlikely to resolve separate L$_3$ peaks, and this was the case for species in Table \ref{tab:ledge}. We consequently averaged the two low energy eigenvalues into a composite L$_3$ value for Table \ref{tab:ledge} (as the energy differences between those eigenvalues were $< 0.1$ eV). However, the full model with three separate peaks may prove necessary in some cases. 
\color{black}
\bibliography{references}
\end{document}